\documentclass[structabstract]{aa}
\usepackage{graphicx}
\usepackage{txfonts}
\usepackage{natbib}
\bibpunct{(}{)}{;}{a}{}{,}
\begin{document}
   \title{Evolution of the optical Tully--Fisher relation up to z=1.3}

   \author{M. Fern\'andez Lorenzo\inst{1}, J. Cepa\inst{1}$^,$\inst{2}, A. Bongiovanni\inst{1}, H. Casta\~neda\inst{1}, A.M. P\'erez Garc\'ia\inst{1}, M.A. Lara-L\'opez\inst{1}, M. Povi\'c\inst{1}, \and  M. S\'anchez-Portal\inst{3}
          }

   \institute{Instituto de Astrof\'isica de Canarias (IAC),
              C/ V\'ia L\'actea S/N, 38204 La Laguna, Spain\
%              \email{mfernan@iac.es}
   \and
     Departamento de Astrof\'isica, Universidad de La Laguna, Spain
         \and
             INSA--ESAC, Madrid, Spain
            }

% \date{Received.....; accepted..... }

% \abstract{}{}{}{}{} 
% 5 {} token are mandatory

  \abstract
   {The study of the evolution of the Tully--Fisher relation has been controversial in the past years. The main difficulty is in determining the required parameters for intermediate and high redshift galaxies, given the cosmological corrections and biases involved.}
   {This work aims to identify the main problems of the study of the Tully--Fisher relation at high redshift using optical emission lines, in order to draw conclusions about the possible evolution of this relation in the B, R, and I--bands.}
   {With this aim, the rotational velocities obtained from the widths of different optical lines using DEEP2 spectra are compared. This method allows reaching higher redshifts against the rotation curve one, due to spatial resolution limitations. Morphology has been determined via HST images, using and comparing different methodologies. Instrumental magnitudes are then corrected for K and extinction and the absolute magnitudes derived for the concordance cosmological model. Finally, the optical Tully--Fisher relations in B, R, and I--bands at different redshifts up to $z = 1.3$ are derived.}
   {Although most studies (this one included) find evidence of evolution, the results are not conclusive enough, since the possible luminosity evolution is within the scattering of the relation, and the evolution in slope is difficult to determine because at high redshift only the brightest galaxies can be measured. Nevertheless, our study shows a clear tendency, which is the same for all bands studied, that favours a luminosity evolution where galaxies were brighter in the past for the same rotation velocity. This result also implies that the colour of the Tully--Fisher relation does not change with redshift, supporting the collapse model versus the accretion model of disc galaxy formation.}
   {}

   \titlerunning{Evolution of the optical Tully--Fisher relation up to z=1.3}
   \authorrunning{Fern\'andez Lorenzo et al.}
   \maketitle

%
%________________________________________________________________

\section{Introduction}

The Tully--Fisher relation (hereafter TFR) relates the luminosity to the maximum rotational velocity of spiral galaxies \citep{1977A&A....54..661T}. This relation is valid for all types of spirals. For example, \citet{1995MNRAS.273L..35Z} obtained the same TFR for high and low surface--brightness galaxies, while \citet{2003ApJ...594..208C} came to the same conclusion for barred spirals. 

Then, the TFR represents a relation between fundamental galaxy parameters, such as its total mass, and the mass locked in stars. This valuable information about the relation between dark and luminosity matter \citep{1995MNRAS.273L..35Z} has important implications for the study of the evolution of disc galaxies. Its slope is an input parameter for galaxy formation models, and its absolute luminosity level provides constraints for the mass--to--light ratios and the initial mass functions \citep{2007MNRAS.375..913P}. Finally, the TFR is an important distance estimator that has been traditionally used for measuring the Hubble constant H$_0$ \citep[e.g.][]{2000ApJ..533..744}. 

For these reasons, the study of the possible evolution of the TFR by comparing what has been obtained from galaxies at high redshift with their local counterpart, has outstanding importance and far reaching consequences for determining fundamental cosmological parameters in the study of structure formation, and in the evolution of disc galaxies \citep{2000ApJ...538..477N}. 

In the past years, some groups have measured the TFR at different redshifts. Pioneer studies, most of them in the B--band, seem to show evolution, although there was no consensus on whether this evolution was on slope, zero point, or both. \citet{2002ApJ...564L..69Z} find a shallower relation at high redshift (z$=$1) than what has been measured in local samples. In their study, they claim that less massive galaxies were 1 or 2 magnitudes brighter in the past (at fixed rotational velocity), while the most massive would follow the local relation. \citet{2004A&A...420...97B} find the same evolution on slope and evidence for a luminosity evolution with a look--back time of ${\Delta}M_B \approx -1$ magnitudes at redshift z$=$1. However, other groups, such as \citet{1997MNRAS.285..779R} or \citet{2006MNRAS.366..308B}, find luminosity evolution but no slope change. Moreover, these authors find differences in luminosity evolutions ranging from 0.2 to 2 magnitudes. 

Since optical bands, especially in the blue, are more affected by extinction, infrared TFR have been studied since very early on \citep{aaronson79}. When the infrared TFR was recently applied to evolutionary studies, \citet{2005ApJ...628..160C} and \citet{2006A&A...455..107F} did not find a K--band TFR evolution. However, in a recent work, \citet{2008A&A...484..173P} do find evolution in the sense that galaxies had been fainter in the past, a result opposite to what is found in the optical bands \citep[for example,][]{2004A&A...420...97B}. These results suggest that the TFR could be dependent on the observed photometric band. An alternative method for studying the TFR has been derived by \citet{2001ApJ...550..212B} using the stellar--mass obtained from the mass--to--light ratio. These authors claim that it has the advantage of being band--independent. The principal disadvantages are, however, its larger scatter, and an initial mass function (IMF) must be assumed. Even so, some studies have used this method \citep[e.g.,][]{2005ApJ...628..160C}, finding no evolution in the stellar--mass TFR. Nevertheless, from the combination of the rotational velocity and velocity dispersion, \citet{2007ApJ...660L..35K} reduced the stellar mass TFR scatter, concluding that there is no evolution up to z=1.2. In summary, all these previous works indicate that the issue of TFR evolution if far from being settled.

While there seems to be a consensus on the absence of evolution in the stellar--mass TFR, the results obtained from intermediate--redshift optical studies (0.3$<$z$<$0.7) show discrepancies that could be attributed to observational biases. Some works have attempted to pinpoint these limitations, such as the one by \citet{2001AJ....121..625B}, who conclude that the possible explanations for the TFR outliers (i.e. objects with substantial deviations from the local TFR) could be gas infall, non--uniform or truncated emission, and disc--enhanced star formation. In addition, \citet{2004AJ....127.2694K} find that the local TFR outliers associated with kinematic anomalies occupy the same region of TFR parameter space as the galaxies responsible for TFR slope evolution found in some high--z studies \citep[e.g.][]{2002ApJ...564L..69Z}, which tend to use morphology--blind selection criteria. Also, \citet{2004A&A...420...97B}, and \citet{2006MNRAS.366..144N} conclude that the evolution observed is strongly dependent on selection criteria. Moreover, in a recent study, \citet{2007IAUS..235....3V} does not find evolution up to $z=1.1$, when discarding ellipticals and face--on spirals from her study, instead of applying the more commonly used criteria of selecting galaxies of intense lines or by their inclination angles. Therefore, object selection is of outstanding importance for drawing conclusive results on the evolution of the TFR. However, when extending the study of the TFR to higher z, it is difficult to segregate interactive and anomalous galaxies, which are frequently included in the sample.

The rotation velocity measurement is another challenge in TFR studies. At high z, the spatial resolution required to resolve the rotation curve is comparable to the atmospheric seeing. For example, at $z=1$ the Galaxy would have an apparent diameter of about 3 arcseconds. In these cases, integrated velocities are the most convenient way to estimate the velocity field of disc galaxies. From the relation between the velocity field obtained from line widths and from rotation curves, \citet{2007IAUS..235....3V} concludes that both methods are equivalent. 

There are other types of errors, such as inclination angle determination or extinction correction, that tend to shift a galaxy along the TFR rather than perpendicular to it \citep{2007IAUS..235....3V,2006A&A...450...25V}; as a consequence, they do not introduce fake evolution in the TFR at intermediate or high redshift.

In this work, a study of the rotation velocity from optical emission lines has been carried out using different lines, with the aim of reliably increasing the redshift range, reaching up to $z=1.6$ in the optical and even more in the infrared. We find that, although all optical lines can be used to calculate the rotation velocity, the line fitting technique and the signal--to--noise ratio play important roles in its proper determination at high redshift. In addition, we used three different techniques for morphological classification to look for the best method for discarding E/S0 type and interactive galaxies from our sample. For the first time, B, R, and I band TFRs are estimated using a significant sample of galaxies up to $z=1.3$. With the study of the TFR evolution in several optical bands, its possible colour differences versus redshift could provide clues to the mechanism responsible for this evolution. 

This paper is organised as follows. In Sect. 2, the DEEP2 spectroscopy, a description of the morphological analysis, and sample selection criteria are provided. The study of the optical lines and derivation of the rotation velocity and luminosities are explained in Sect. 3. Finally the results are given in Sect. 4, whereas the last two sections provide the discussion of the results and the conclusions. Throughout this article, the concordance cosmology with ${\Omega}_{\lambda}=0.7$, ${\Omega}_m=0.3$ and $\rm H_0=70 \rm \ km \rm \ s^{-1} \rm \ Mpc^{-1}$ is assumed. All magnitudes are in the AB zero--point system.

\section{Data}

The sample consists of galaxies in a 16$^\prime\times$16$^\prime$ field in the Groth Strip Survey (GSS) sky region, which will be observed by the OTELO project \citep[OSIRIS Tunable Emission Line Object Survey,][]{2008A&A...490....1C}. This project will survey emission--line objects using the OSIRIS \citep{2005RMxAC..24...82C} tunable filters in the wavelength intervals defined by windows through the OH emission line forest. OSIRIS is the optical Day One instrument of the GTC 10.4m telescope. 

The baseline for spectroscopy target pre--selection were the galaxies for which DEEP2 spectra (Data Release 3, DR3) in this field were available. In the redshift range 0.1$<$z$<$1.3, there are 1200 galaxies in the DEEP2 DR3 sample with a B--band limiting magnitude of 24.

\subsection{The DEEP2 Database}

The DEEP2 project \citep{2003SPIE.4834..161D,2007ApJ...660L...1D} is a survey using the Keck telescopes to study the distant Universe. The survey targeted a total of ${\sim}10.000$ distant galaxies in the redshift range 0$<$z$<$1.4, using the DEIMOS multi--object spectrograph \citep{2003SPIE.4841.1657F}. The grating used was the 1200 l/mm one, covering a spectral range of 6500--9100 {\AA} with a dispersion of 0.33 {\AA}/px, equivalent to a resolution $R=\lambda/\Delta\lambda$ 4000. The DR3 constitutes the third public data release for this survey and includes redshifts plus B, R, and I band photometry, in addition to 1D and 2D spectra. The photometric data were taken with the CFH12K mosaic camera \citep{2001ASPC..232..398C}, which is installed on the 3.6--meter Canada--France--Hawaii Telescope (CFHT). In the DEEP2 photometric catalogue (Data Release 1, DR1) \citep{2004ApJ...617..765C} all magnitudes were corrected for reddening using \citet{1998ApJ...500..525S} dust maps. Magnitude errors from sky noise and redshift are available too.

As already pointed out, since resolved rotation curves are difficult to obtain at high redshift due to the limited spatial resolution, it has been necessary to use the integrated spectra obtained by fitting line profiles to the 1D spectra provided by DR3 \citep{1986PASP...98..609H}.

\subsection{Morphology}

As stated in Sect. 1, morphological classification plays an important role in TFR determinations. Studies with no morphological selection criteria include objects such as blue compact galaxies, emission--line S0, or galaxies with peculiar morphologies, together with normal spirals.

In the DEEP2 catalogue, the objects are already divided in galaxies, stars, or AGNs, according to a best--fitting template technique. However, a morphological classification of the objects thus identified as galaxies is still required. As a preliminary procedure and with the aim of segregating E/S0 galaxies from non--E/S0 in our sample, we compared the DEEP2 spectra with a synthetic, early--type galaxy spectrum. The chosen template corresponds to a 5 Gyr--old stellar population with a stellar metallicity of 0.02 Z$_{\odot}$ \citep[assuming an IMF from][]{2003PASP..115..763C}, which is part of the \citet{2003MNRAS.344.1000B} collection of galaxy templates used by \citet{2003AAS...202.5102T} in the analysis of SDSS galaxy spectra. Then, the r.m.s. of the differences between the template and every object were calculated to determine whether they matched each other or not. The result was that about one--third of the spectra could be associated with E/S0 galaxies. Nevertheless, when the HST images of a statistically significant subsample of each group were visually analysed, the classification mismatch for the non--E/S0 group was $\sim$3\%, whereas the one associated with the group of E/S0 galaxies was significantly worse. This result forced us to proceed to a visual classification of every galaxy using HST images, as the most reliable procedure for an accurate determination of morphological types. The HST images used are part of AEGIS survey \citep{2007ApJ...660L...1D} and were obtained with the ACS camera. There are 809 galaxies in our field that have a DEEP2 spectrum, and V and I--band HST images. In our morphological classification we divided the objects into five groups: elliptical/S0 (8\%), spirals (69\%), irregulars (6\%), interactive (6\%), and unknown (11\%). To classify our visually unknown objects, GIM2D \citep{1998ASPC..145..108S} was used. In addition, to somehow estimate the reliability of the GIM2D classification, GIM2D parameters were also determined for other types. For the galaxies visually classified as spirals, the S\'ersic index was obtained for a random subsample of 170 galaxies. An apparent classification mismatch was found for 40 objects ($\sim$24$\%$). However, 25 of them were clearly spirals, even though the S\'ersic index calculated was greater than 2.5. The S\'ersic index was also determined for a random subsample of galaxies visually classified as ellipticals. In this case, $\sim$60$\%$ had a S\'ersic index lower than 2.5, although half of them were clearly ellipticals. These results show that our visual classification has clear advantages over that obtained using GIM2D.

\begin{figure}
   \centering
   \includegraphics[angle=90,width=7.9cm]{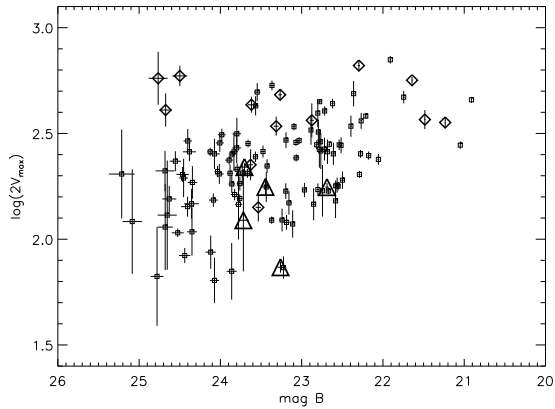}
      \caption{Maximum rotational velocity vs. instrumental magnitude, in the redshift interval 0.3$<$z$<$0.8. Diamonds are galaxies with inclination angle lower than 25$^{\circ}$. Squares represent galaxies with inclination angles between 25$^{\circ}$ and 75$^{\circ}$. Triangles are galaxies with inclination angles between than 75$^{\circ}$ and 80$^{\circ}$ (the limit in our sample).}
         \label{Fig1}
   \end{figure}

\subsection{Sample selection}

In some studies of the TFR, the sample is selected according to the emission line observed. For example, \citet{2001AJ....121..625B} considered only galaxies with H$_{\alpha}$ emission lines, while \citet{2006A&A...455..107F} studied only galaxies with [OII]$\lambda\lambda$ 3727\AA\ emission. In the present work, all conspicuous optical emission lines will be used to expand the redshift range to the maximum possible. 

The inclination angle (i) is an usual selection criterion as well. For instance, \citet{2006MNRAS.366..144N} study galaxies with i$>45^{\circ}$, and \citet{2007ApJ...660L..35K} choose spirals with 30$^{\circ}<$i$<$70$^{\circ}$. In the first case, this restriction can include galaxies with large inclination errors (almost edge--on galaxies), and in both cases, the limits seem too restrictive. With the aim of avoiding selection criteria that can bias the results, morphology and inclination selection will be carefully studied. The inclination angle can be obtained from the major to minor axes ratio as found in the HST catalogue. In order to obtain the errors, we used {\tt sextractor} \citep{1996A&AS..117..393B} in the summed V+I images and compared the results. We estimated the mean error to be ${\sim}$ 2.5 degrees, in agreement with \citet{2006A&A...455..107F}. In Fig. 1, galaxies with three different inclination ranges have been represented. No correlation between magnitude and rotation velocity can be observed for inclinations lower than 25$^{\circ}$. Also, galaxies almost edge--on will be more affected for extinction and the error in inclination angle will be larger.

In Fig. 1, the galaxies with inclination angle up to 75$^{\circ}$ seem to follow a TFR, and in our sample all galaxies have inclination angles lower than 80$^{\circ}$. For these reasons, only galaxies with inclinations between 25$^{\circ}$ and 80$^{\circ}$ will be considered.

Finally, only normal spiral galaxies follow the TFR. Then, a first selection of spirals was done using our visual classification. All galaxies belonging to E/S0 and interactive groups were discarded. The irregular galaxies were, nevertheless, included, since they cannot be distinguished from spiral galaxies at high redshift. Either way, the number of irregular galaxies is not significant, and they affect low and high redshift regimes as well. There are a few objects visually classificated as unknown and without S\'ersic index that were eliminated because their morphology was not known. However, it is very difficult to do a reliable classification at high redshift, and then peculiar galaxies with kinematic anomalies could be included in our study.

Figure 2 shows that the redshift distribution of all DEEP2 galaxies in our field, and that of our final sample, after morphology and inclination selection, are very similar.  

 \begin{figure}
   \centering
   \includegraphics[width=8.3cm]{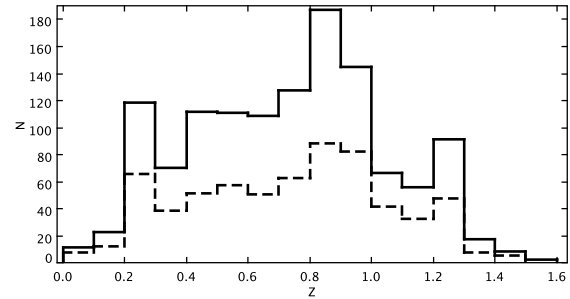}
      \caption{Redshift distribution of all DEEP2 galaxies in our field (solid line) and of galaxies selected according to our morphology classification and inclination criteria (dashed line).}
         \label{Fig2}
   \end{figure}

\begin{figure*}
\centering
\includegraphics[scale=0.46]{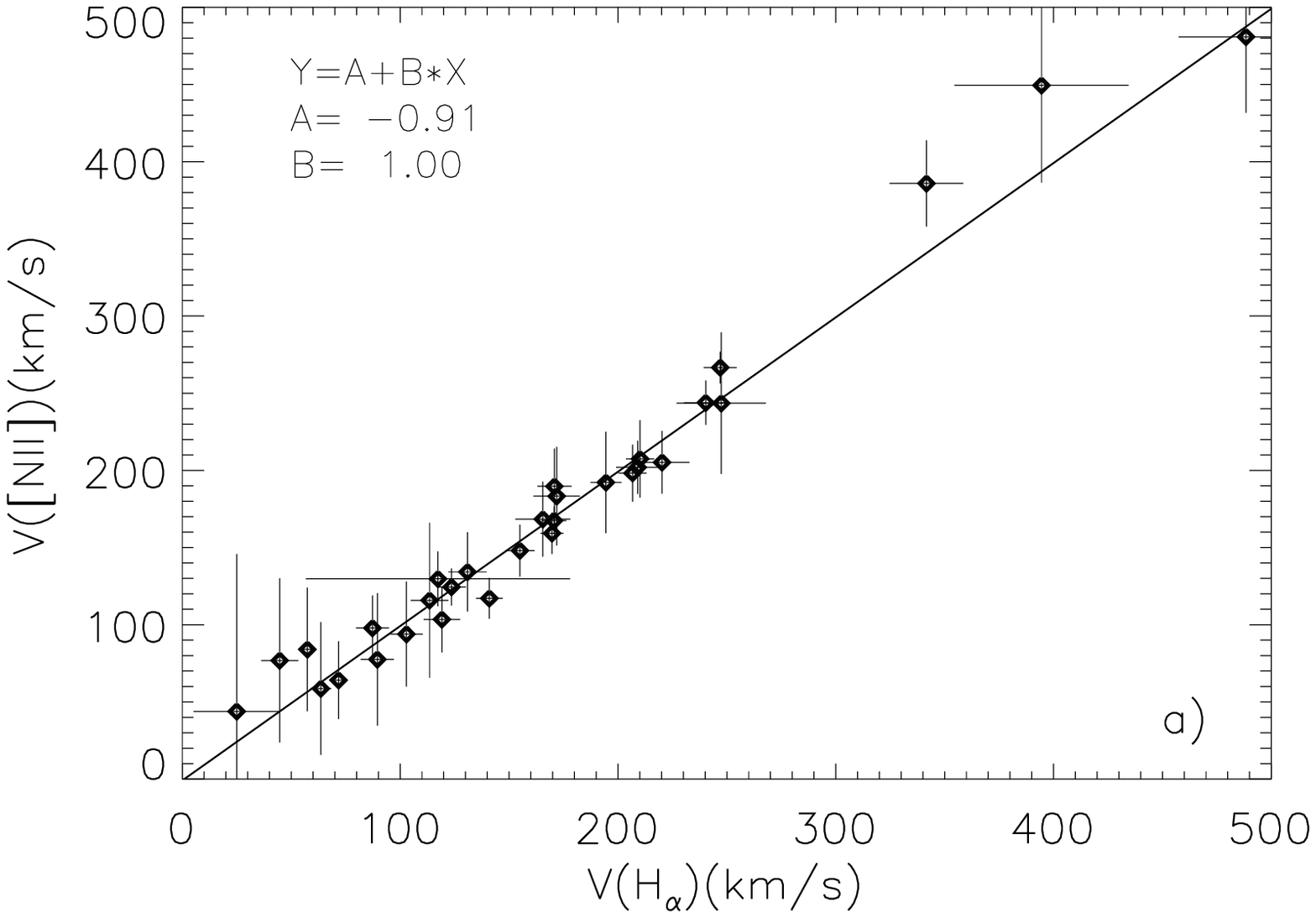}
\includegraphics[scale=0.46]{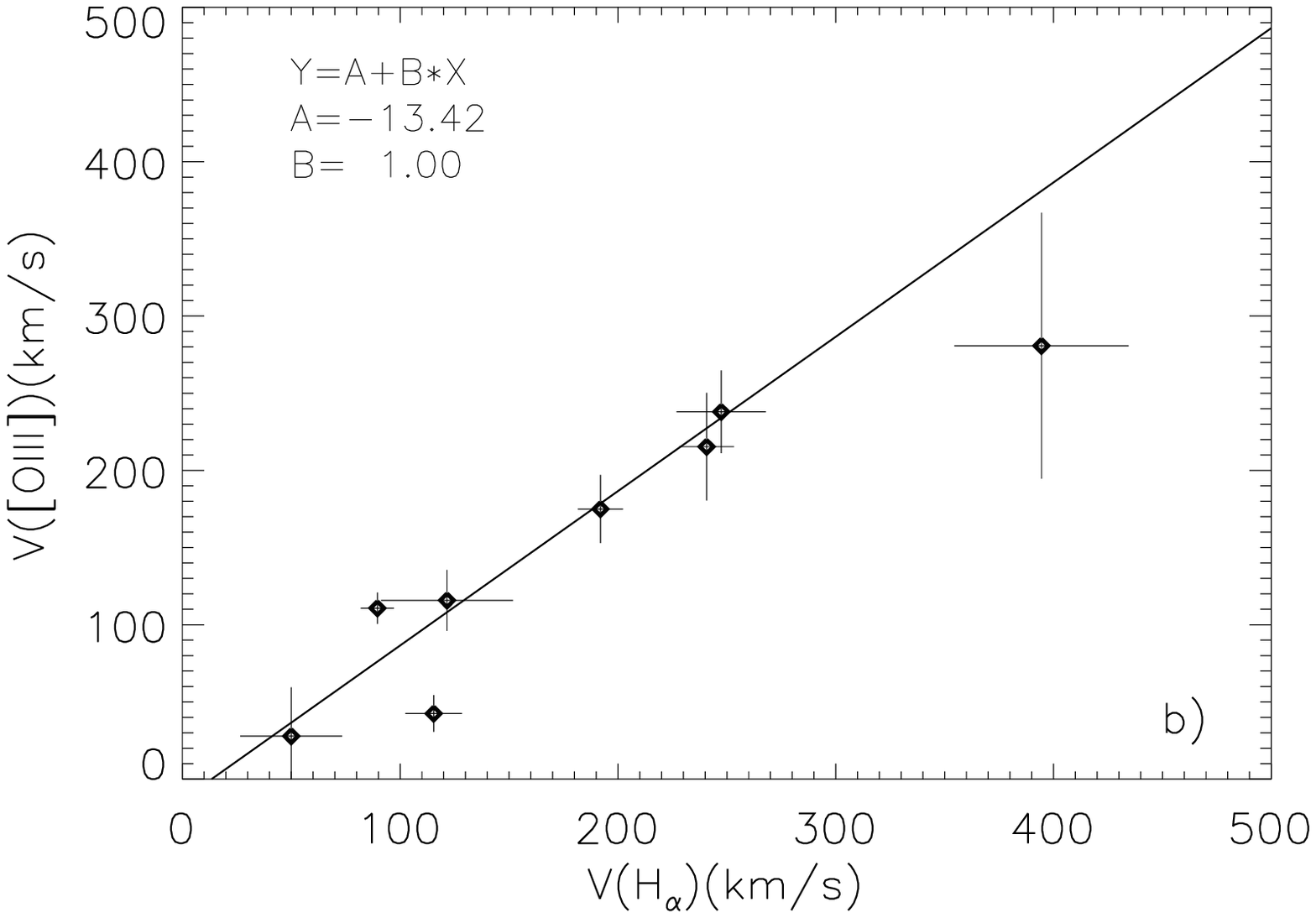}
\includegraphics[scale=0.46]{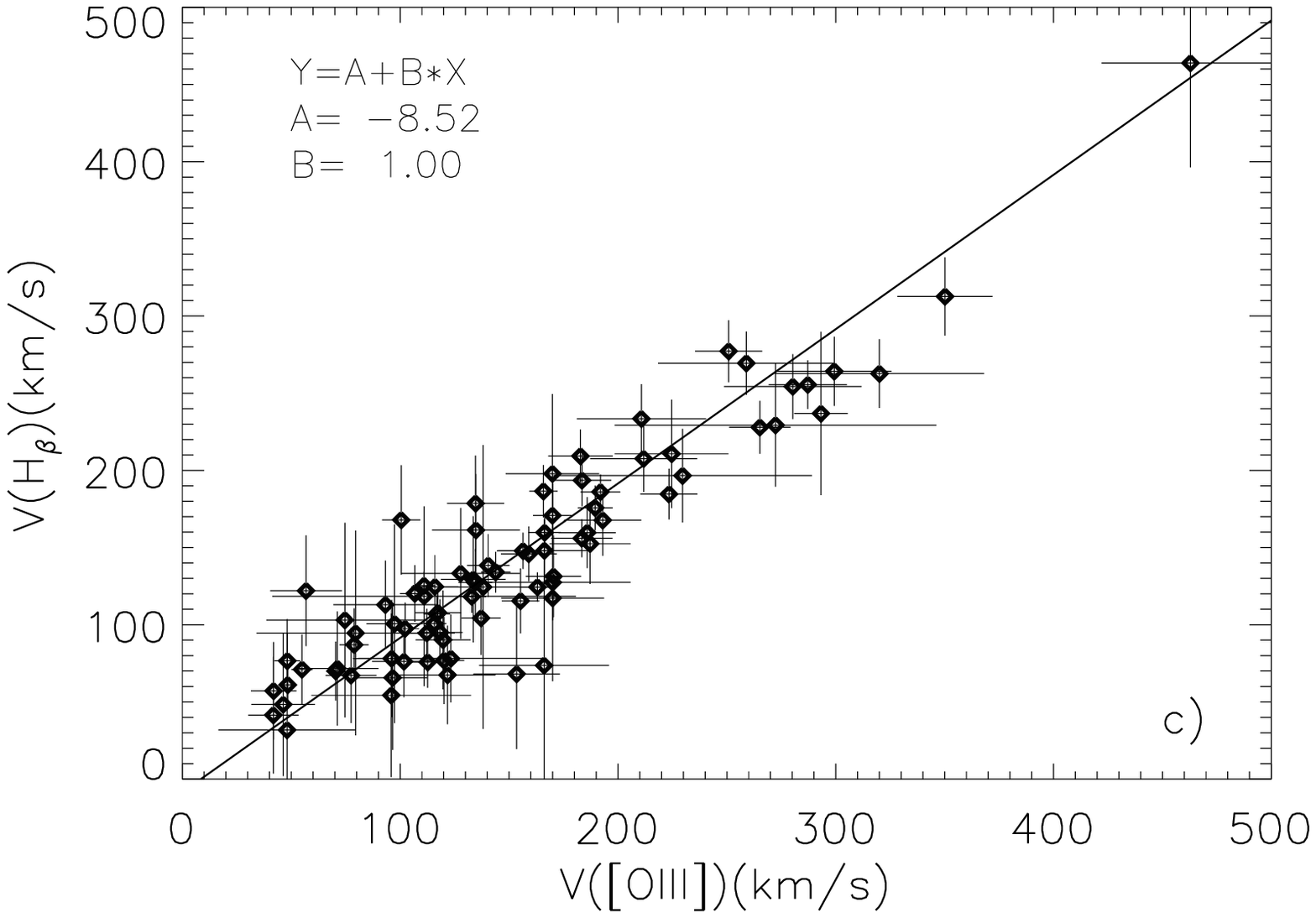}
\includegraphics[scale=0.46]{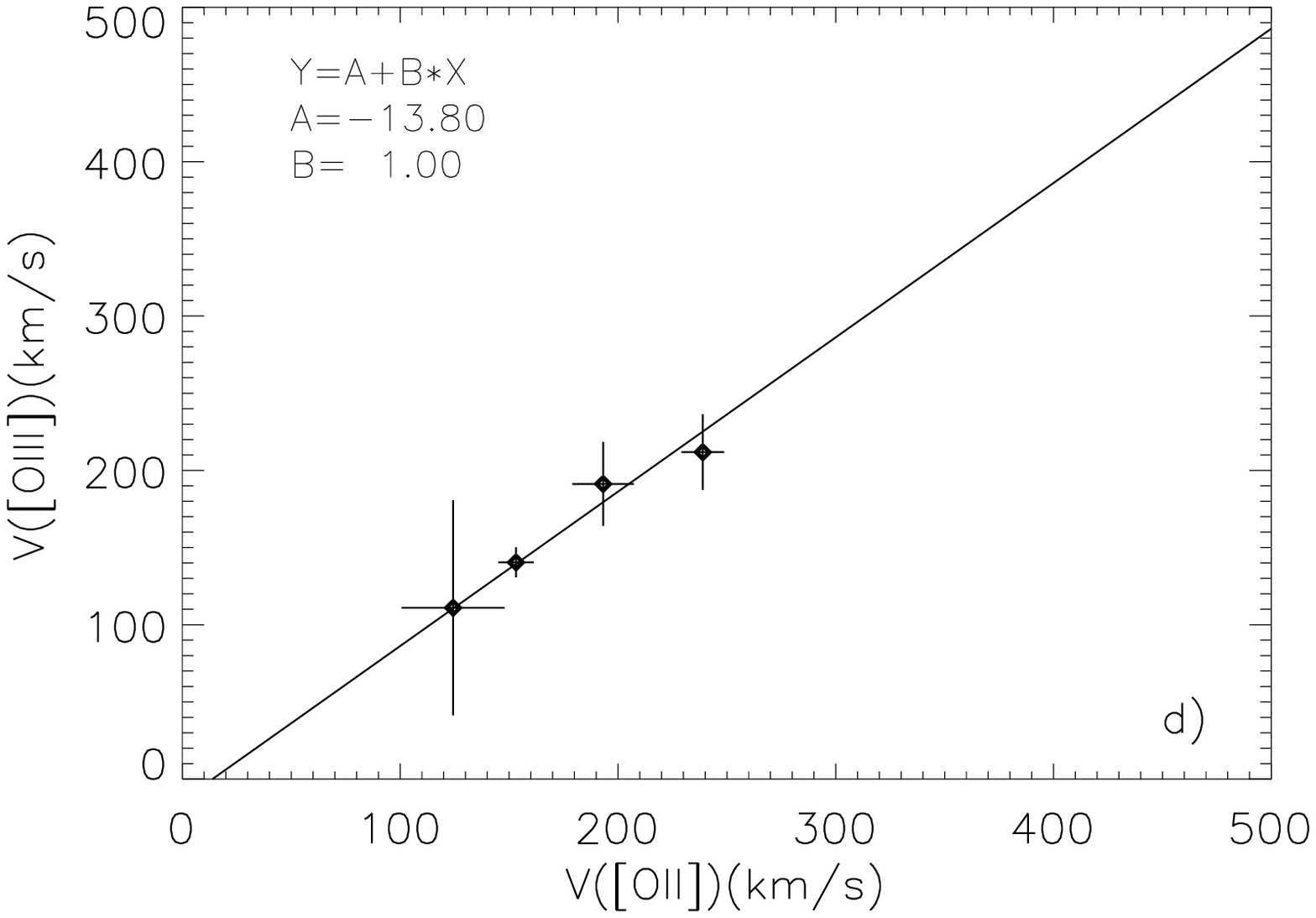}
\caption{Comparison of the velocity widths from five optical lines. In each panel, the solid line shows the least--square fits with slope 1. All zero--points are within the errors. a) Velocity obtained from [NII]$\lambda$6583\AA\ and H$_{\alpha}$, b) comparison between [OIII]$\lambda$5007\AA\ and H$_{\alpha}$, c) velocity from H$_{\beta}$ versus [OIII]$\lambda$5007\AA, d) comparison between [OIII]$\lambda$5007\AA\ and [OII]$\lambda\lambda$3727\AA\ velocities.}
\end{figure*}

\section{Deriving Tully--Fisher relations}

The final sample of galaxies with redshifts and B, R, and I magnitudes, classified as spirals, and with inclinations in the range between 25$^{\circ}$ and 80$^{\circ}$ is composed of 612 galaxies. The next step is determining inclination and redshift--corrected magnitudes and rotation velocities. The derivation of the Tully--Fisher relation also requires determining absolute B, R, and I--band magnitudes. They have been obtained by K--correcting B, R, and I--band apparent magnitudes using \citet{2007AJ....133..734B}, by applying a concordance cosmology for determining the luminosity distance from the redshifts, and by correcting for intrinsic extinction \citep{1998AJ....115.2264T}.

\subsection{Rotation velocity estimation}

The local TFR has been historically obtained from radio measurements, specifically from 21cm line widths. The HI emission spreads further than optical emission, so that optical and radio measurements can be compared, as long as both extend up to the flat region or to the maximum of the rotation curve. Moreover, the existence of three types of rotation curves, depending on the relation between the maximum velocity and the velocity of the flat region \citep{2001ApJ...563..694V}, can complicate the comparison. Also, turbulence plays a role in the optical versus radio velocity width determination. For example, \citet{1992ApJS...81..413M} compared the projected rotation velocity measured from H$_{\alpha}$ rotation curves with the velocity measured from integrated HI profiles at the 50\% level. They obtain a difference of 10 km/s, which is attributed to HI widths that measure rotational velocity plus internal galaxy turbulence. This turbulence is important in the most external region of the galaxy where the gravity is lower. The optical emission does not spread until this region, so the turbulence is negligible against the rotational velocity determined using these lines:

 \begin{equation}
 2V_{max} = \frac{{\Delta}{\lambda} \ {\rm c}}{\lambda_{0}\sin{\rm (i)} \ (1+z)} \ .
 \end{equation}

\begin{figure*}
\centering
\includegraphics[scale=0.46]{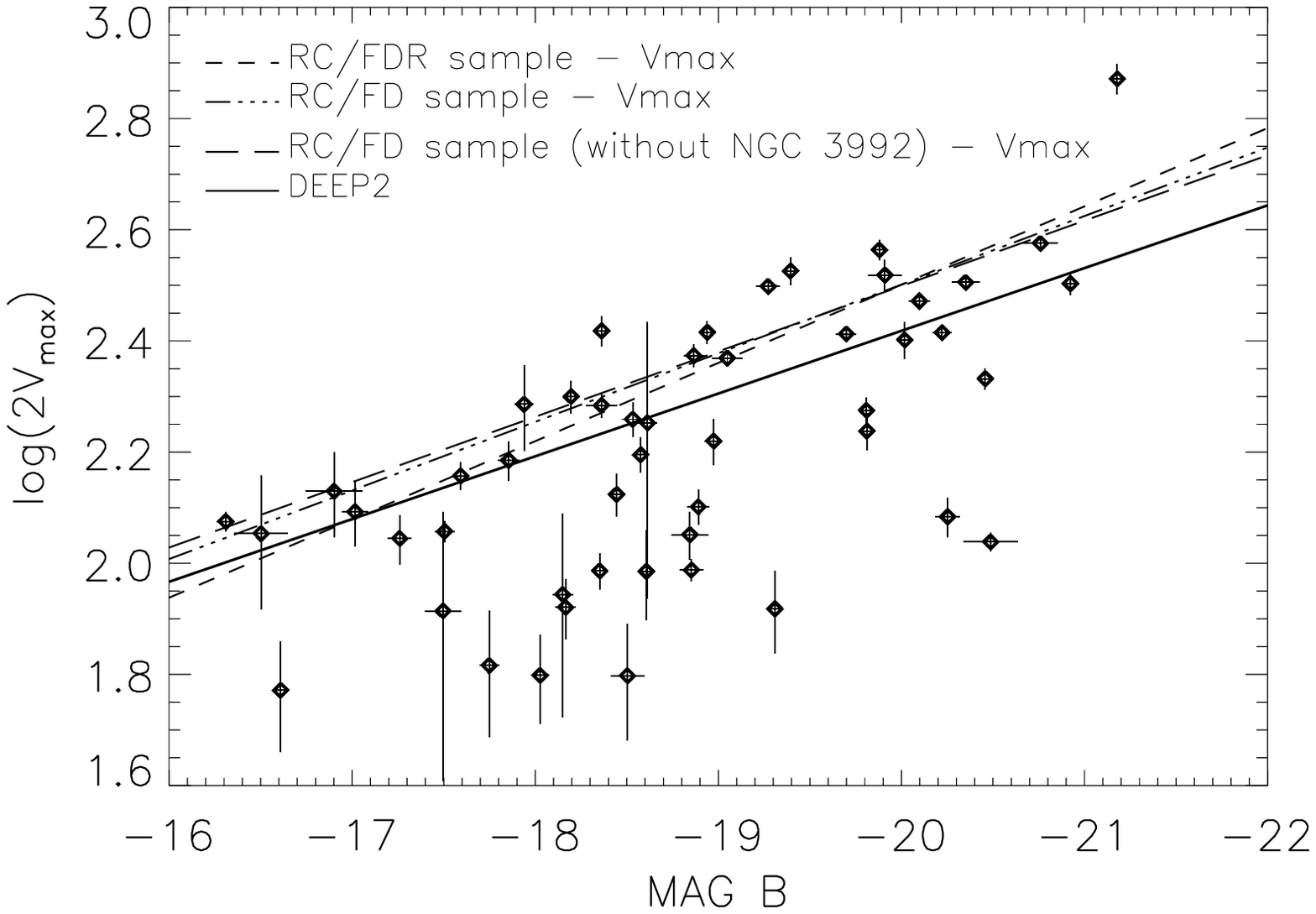}
\includegraphics[scale=0.46]{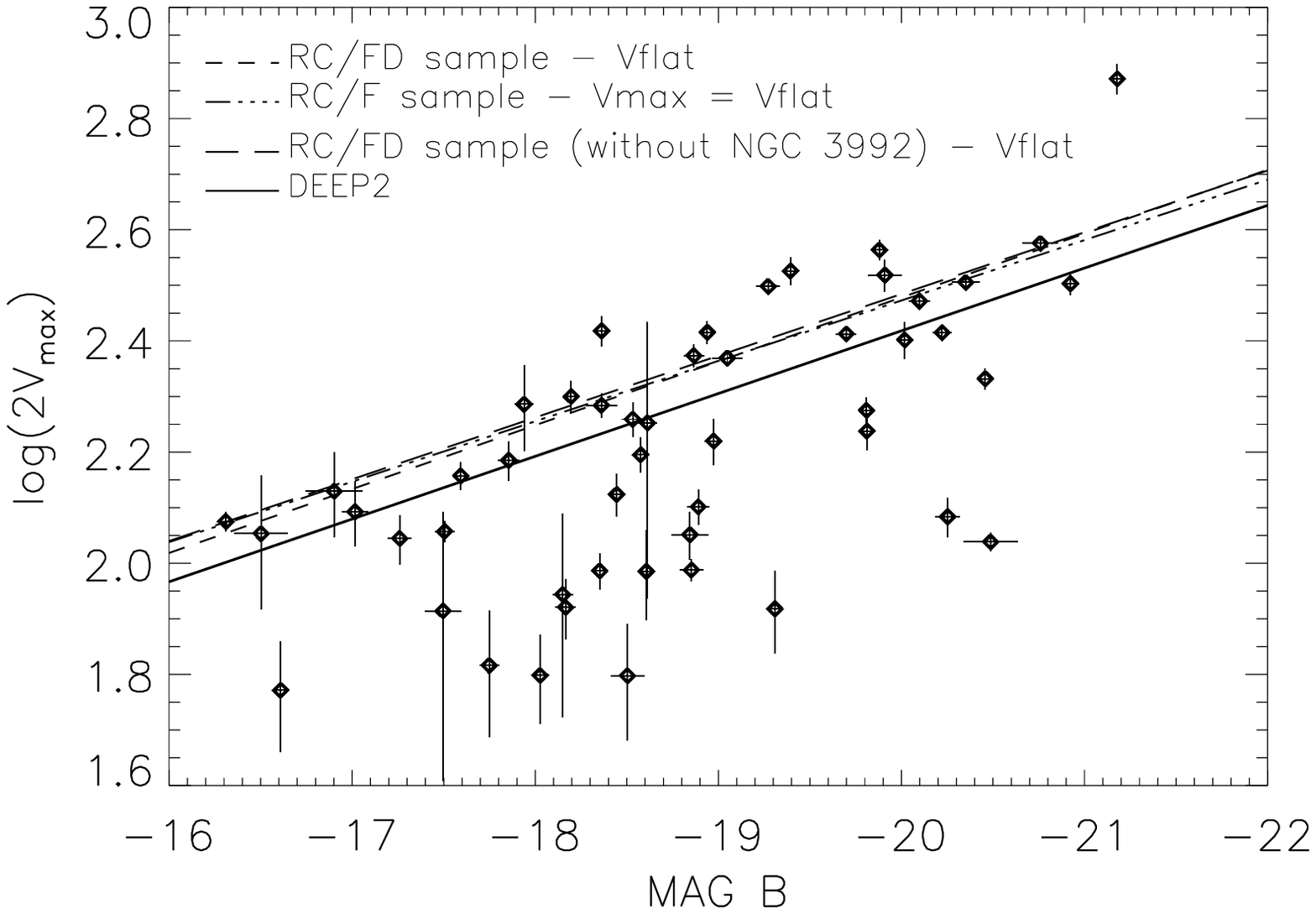}
\caption{Comparison between local TFR obtained from DEEP2 data (0.1$<$z$<$0.3) in B--band, and \citet{2001ApJ...563..694V} local TFRs. The solid line represents the linear fit to the DEEP2 data (See text for definitions of the subsamples).}
\end{figure*}

To measure rotation velocity, we used equation (1), where i is the inclination angle, ${\lambda}_0$ the line central wavelength at z=0, and ${\Delta}{\lambda}$ the line width at 20\% of peak intensity. We determined ${\Delta}{\lambda}$ with the Gaussian fitting routines in the Starlink package {\tt dipso} \citep{1996sun50}, which calculate the FWHM (full--width half--maximum). Twenty percent of the peak intensity is related with the FWHM as $W_{20}=1.524 \ {\rm FWHM}$. Finally, we corrected for the instrumental width using the expression

\begin{equation}
 {\Delta}{\lambda} = \sqrt{{\Delta}{\lambda}_{obs}^2 - {\Delta}{\lambda}_{ins}^2} \ .
 \end{equation}

The rotational velocities obtained from the widths of different optical lines are compared in Fig. 3. Although all velocities are remarkably similar, a larger dispersion in the H$_{\beta}$ vs [OIII]$\lambda$5007\AA\ can be seen. The reason is that H$_{\beta}$ has low signal--to--noise in most spectra and is heavily affected by stellar absorption, so this line has not been used in the study. The fit is quite good in the last panel, where [OIII]$\lambda$5007\AA\ is compared versus [OII]$\lambda\lambda$3727\AA, although only four objects have both lines in common. It is important to point out that a fifth galaxy with [OII]$\lambda\lambda$3727\AA\ velocity, 50 km/s higher than that obtained using [OIII]$\lambda$5007\AA, was discarded because its spectrum shows that the object was an AGN or another active object. This leaves a final sample of 344 galaxies for which rotational velocities were calculated. 

\subsection{Absolute magnitudes}

The absolute magnitudes were obtained after correcting for Galactic reddening, K--correction, and extinction. DEEP2 magnitudes are provided already corrected for Galactic reddening based on \citet{1998ApJ...500..525S} dust maps.
 
The K--correction was performed by adopting the \citet{2007AJ....133..734B} procedure (code version {\tt v4\_1\_4}), which provides the K--correction from the B, R, and I--band photometry. The reliability of the K--corrections thus obtained depends mainly on the errors of DEEP2 photometric catalogue, which are lower than 0.2 mag. After K--correction, absolute magnitudes and its inverse variance were calculated from the luminosity distance corresponding to the measured redshifts, using the concordance cosmology.

Finally, the absolute magnitudes were corrected for intrinsic extinction. This correction is basically dependent on inclination. Several works have studied this effect, providing different corrections. The \citet{1985ApJS...58...67T} and the \citet{1998AJ....115.2264T} methods are the most common corrections used in TFR studies \citep[e.g.][]{2004A&A...420...97B,2007MNRAS.377..806C}. In this work we have adopted the \citet{1998AJ....115.2264T} method so as to compare it with the most recent TFR studies. According to this method, the extinction, $\rm A_{\lambda}$, as a function of inclination, i, in the ${\rm {\lambda}-band}$, is defined to be:\\
\begin{equation}
A_{\lambda}^{i-0}={\gamma}_{\lambda} \ \rm log \ ({\rm a/b})
\end{equation}
with
\begin{equation}
{\gamma}_B=-0.35 \rm (15.31+M_B)
\end{equation}

\begin{equation}
{\gamma}_R=-0.24 \rm (15.91+M_R)
\end{equation}

\begin{equation}
{\gamma}_I=-0.20 \rm (16.61+M_I)
\end{equation}
where a/b is the galaxy major--to--minor axis ratio.

\section{Results}

\begin{table*}
\caption{Best approximation between this work and \citet{2001ApJ...563..694V} parameters obtained in the sample (iv), by fitting $V_{max}$.}             % title of Table
\label{table:1}      % is used to refer this table in the text
\centering                          % used for centering table
\begin{tabular}{c c c c c c c c}
\hline\hline
\multicolumn{1}{c}{}&\multicolumn{5}{c}{\bf This work}&\multicolumn{2}{c}{\bf \citet{2001ApJ...563..694V}}\\
Band & a & b & ${\sigma}_{total}$ & A & B & A & B \\\hline
B & $0.160\pm0.062$ & $-0.113\pm0.003$ & $0.149$ & $1.42\pm0.55$ & $-8.85\pm0.24$ & $1.24\pm0.82$ & $-8.5\pm0.4$ \\
R & $0.183\pm0.057$ & $-0.108\pm0.003$ & $0.144$ & $1.69\pm0.53$ & $-9.26\pm0.26$ & $1.54\pm0.82$ & $-8.9\pm0.4$ \\
I & $0.219\pm0.055$ & $-0.105\pm0.003$ & $0.142$ & $2.09\pm0.53$ & $-9.52\pm0.27$ & $3.05\pm0.83$ & $-9.6\pm0.4$ \\
\hline
\end{tabular}
\end{table*}

\subsection{Local Tully--Fisher relation}

\begin{figure}
\centering
\includegraphics[scale=0.46]{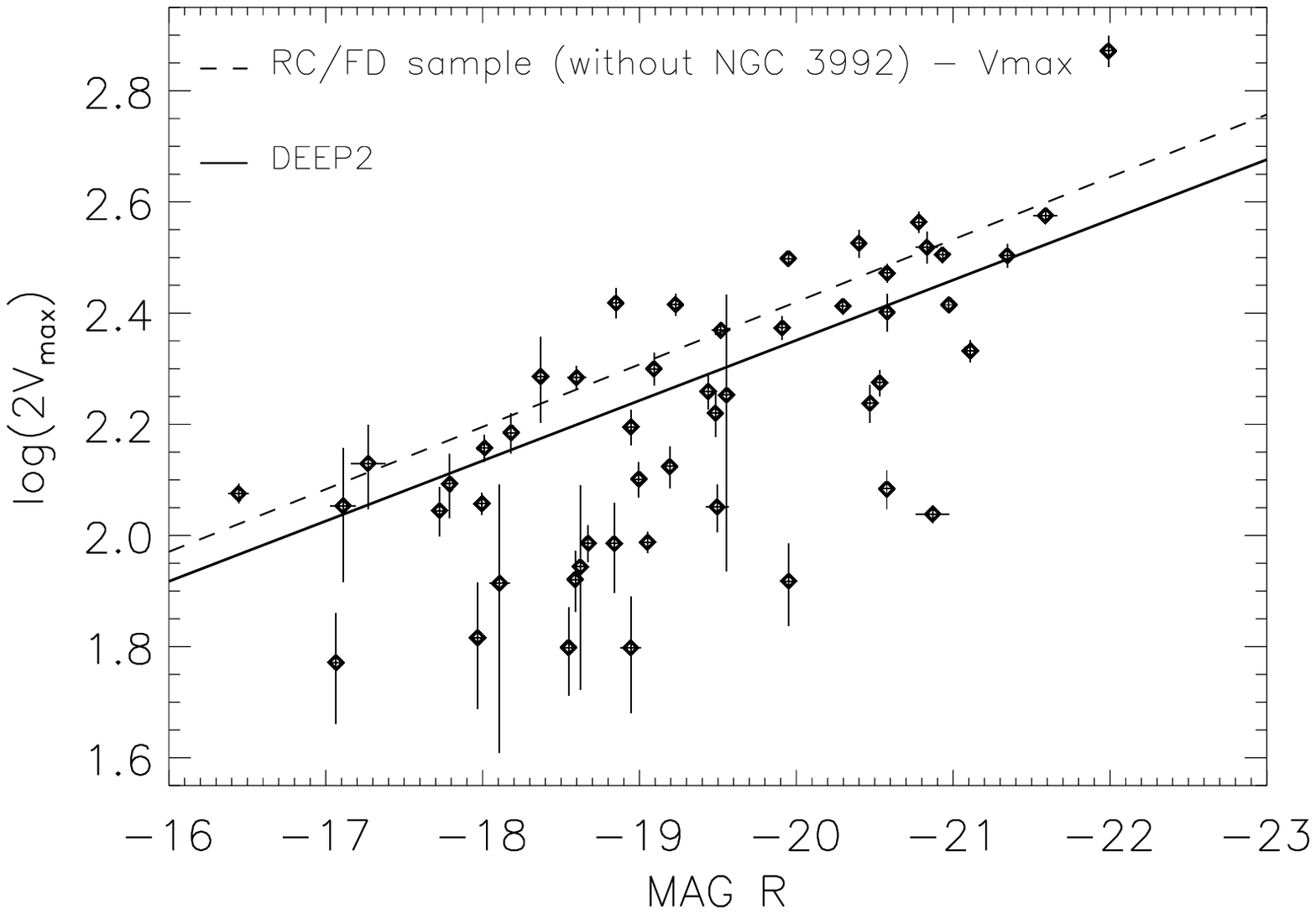}
\includegraphics[scale=0.46]{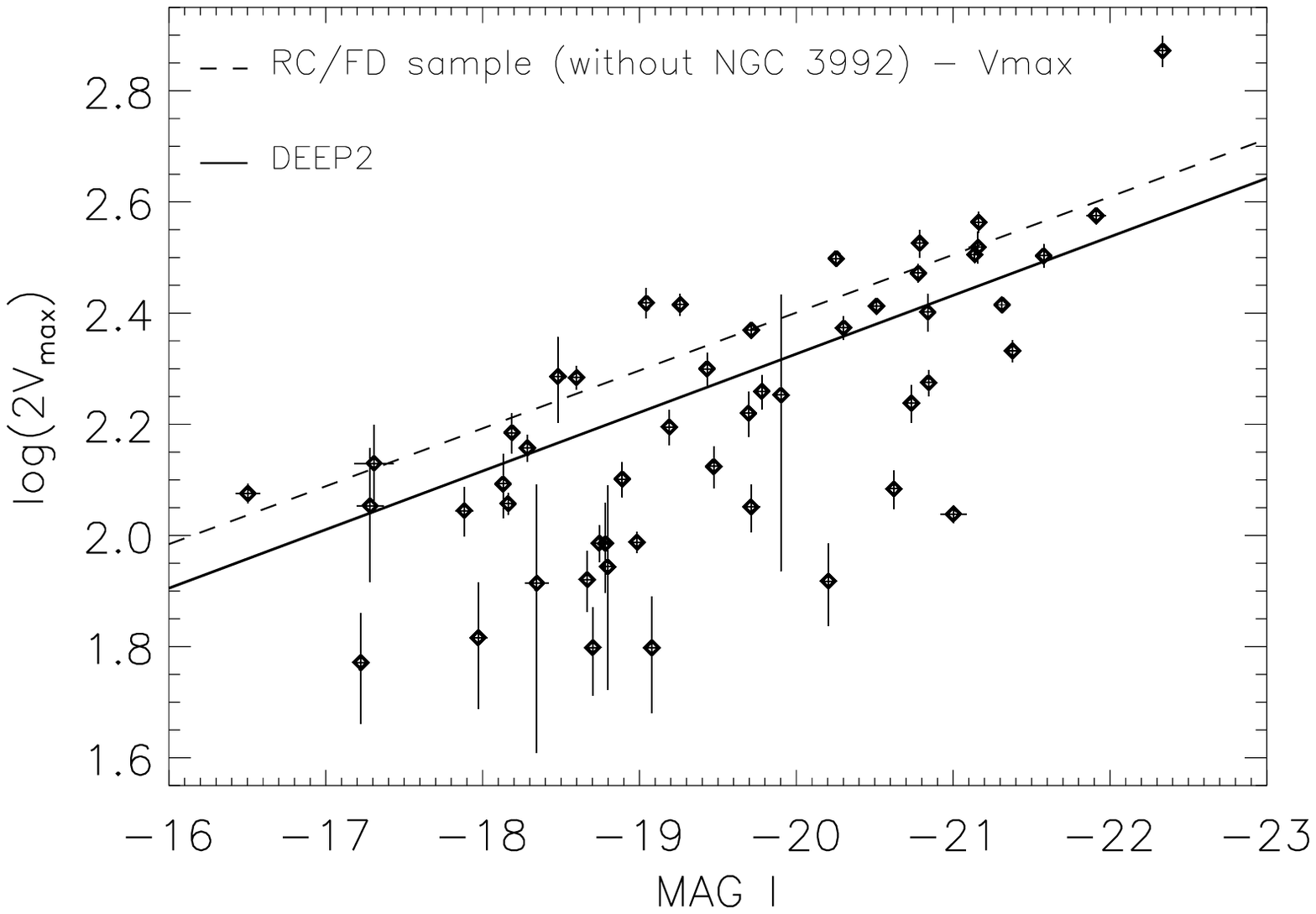}
\caption{Comparison between local TFR obtained from DEEP2 data (0.1$<$z$<$0.3) in R--band (up) and I--band (down), with the \citet{2001ApJ...563..694V} local TFR of the `RC/FD sample (without NGC 3992)'. The latter relation is determined from $V_{max}$, and the galaxies included have $V_{max} \geq V_{flat}$. The solid line represents the linear fit to the DEEP2 data, while the dashed line is the \citet{2001ApJ...563..694V} relation in the respective band.}
\end{figure}

The local TFR in the optical bands have been studied in several works over the years \citep[e.g.][]{2000ApJ..533..744,2001ApJ...563..694V}. The slope and zero--point of the relation are very similar in all of them; however, a small difference in one of these parameters can be definitive for finding evolution. Therefore, a local TFR from the same data and in the same conditions as the high--z study must be established, for a reliable comparison.

The \citet{2001ApJ...563..694V} local relations are widely used in most recent evolution studies of the TFR. For this reason, our local fits are compared with the \citet{2001ApJ...563..694V} relations as a reference. In this paper, different relations are derived as a function of the rotation velocity. The first one uses the parameter $W^i_{R,I}$, defined in \citet{1985ApJS...58...67T}, as the rotation motion parameter derived from the observed HI line width at 20${\%}$ of peak intensity, corrected for random motions, transition from horned to Gaussian profile, and inclination. With optical lines, only the inclination correction is required because we assume Gaussian profiles, and, as stated above, random motions are less important in the optical. 

The other two relations are calculated from the rotation curve. One of them is derived from the maximum rotation velocity, and the other one from the velocity in the flat region of the rotation curve. Since these two velocities might not be the same, different samples are used in the \citet{2001ApJ...563..694V} paper. In the present work, the velocity is calculated from line widths so that it does not correspond exactly with any of them. For this reason, the results of the local TFR obtained from DEEP2 are compared with all \citet{2001ApJ...563..694V} relations derived from different samples:

(i) The `RC/FDR sample' corresponds to all galaxies with resolved rotation curve, including the objects for which the rotation curve is truncated before reaching the flat region (only $ \rm V_{max}$).

(ii) In the `RC/FD sample', the galaxies whose rotation curve is truncated before flat region were eliminated, so it is possible to calculate $ \rm V_{max}$ and $ \rm V_{flat}$.

(iii) In the `RC/F sample', only galaxies with $ \rm V_{max}$=$ \rm V_{flat}$ are included.

(iv) In the `RC/FD sample (without NGC 3992)', this galaxy has been excluded from the `RC/FD sample' because it was responsible for a larger scattering in the relation.

Our local TFR in the B, R, and I--bands was obtained from least--squares fitting to the DEEP2 data points in the redshift range 0.1$<$z$<$0.3. Since no significant change in the stellar populations of field galaxies was observed in this redshift range, we can assume that their TFRs are representative of the local ones. We followed other authors \citep[e.g.][]{2006MNRAS.366..308B,2006MNRAS.366..144N,2007IAUS..235....3V} by adopting $log(2V_{max})$ as the dependent variable in the fit:

\begin{equation}
\log(2V_{max})= \rm a+b \ M_Q
\end{equation}
where $ \rm V_{max}$ is the maximum rotation velocity, a the zero--point of the relation, b the slope, and $M_Q$ the absolute magnitude in the Q--band. This is the so--called inverse TFR, which is less sensitive to luminosity incompleteness bias \citep{1994ApJS...92....1W,1980AJ.....85..801S}.

The best fit to the data was found by minimizing the chi--squared statistics,

\begin{equation}
{\chi}^2= \displaystyle\sum_{i} w_i \left(\log(2V_{max,i})-a-b \ M_{Q,i} \right)^2 \ ,
\end{equation}
where $w_i=1/{\sigma}_i^2$ is the weight applied to each point according to the expression:
\begin{equation}
{\sigma}_i^2={\sigma}_{\log(2V_{max,i})}^2+b^2 \ {\sigma}_{M_{Q,i}}^2 \ .
\end{equation}
Here, ${\sigma}_{\log(2V_{max,i})}$ and ${\sigma}_{M_{Q,i}}$ are the errors derived from the reduction process. These weights are recalculated in each iteration with the new slope, b.

The total scatter in the Tully--Fisher fit is calculated using

\begin{equation}
{\sigma}_{total}^2= \frac{\displaystyle\sum_{i} \ w_i \left( \log(2V_{max,i})-a-b \ M_{Q,i} \right)^2}{\displaystyle\sum_{i} w_i} \ .
\end{equation}
In the \citet{2001ApJ...563..694V} tables, their slope B and zero--point A correspond to the expression:

\begin{equation}
M_Q= \rm A+B \log(V) \ , 
\end{equation}
where log V is either log $W^i_{R,I}$, log $(2V_{max})$ or log $(2V_{flat})$, depending on the case.

Figure 4 shows the comparison of our local DEEP2 TFR in the B--band, with the four \citet{2001ApJ...563..694V} samples. The smaller difference between both relations is obtained using the sample (ii) from $V_{flat}$. However, when considering the three bands all together, the smallest difference is obtained using the sample (iv) by fitting $V_{max}$. All the parameters obtained are within the errors derived by \citet{2001ApJ...563..694V} (see Table 1). Figure 5 shows the comparison between this sample and the DEEP2 TFRs in R and I--bands; therefore, the TFRs obtained from DEEP2 data in the redshift range 0.1$<$z$<$0.3 were adopted as our local TFR, since they were derived in a similar way to the higher redshift galaxies.

\subsection{Tully--Fisher relations in B, R, and I--bands}

With all corrections applied, the TFRs in B, R, and I--bands are obtained, dividing the redshift range 0.1$<$z$<$1.3 into 6 bins, as shown in Fig. 6 for the B--band. At the highest redshift, only the brightest galaxies can be seen, and since the scatter spreads the data points, the fit was performed by setting a constant slope, to study the evolution of the zero--point. In other words, a change in the slope cannot be distinguished from a change in zero--point. The fits were made in the same way as the local fit, but in this case no iterations were needed because the slope was considered invariant.

\begin{figure*}
\centering
\includegraphics[scale=0.46]{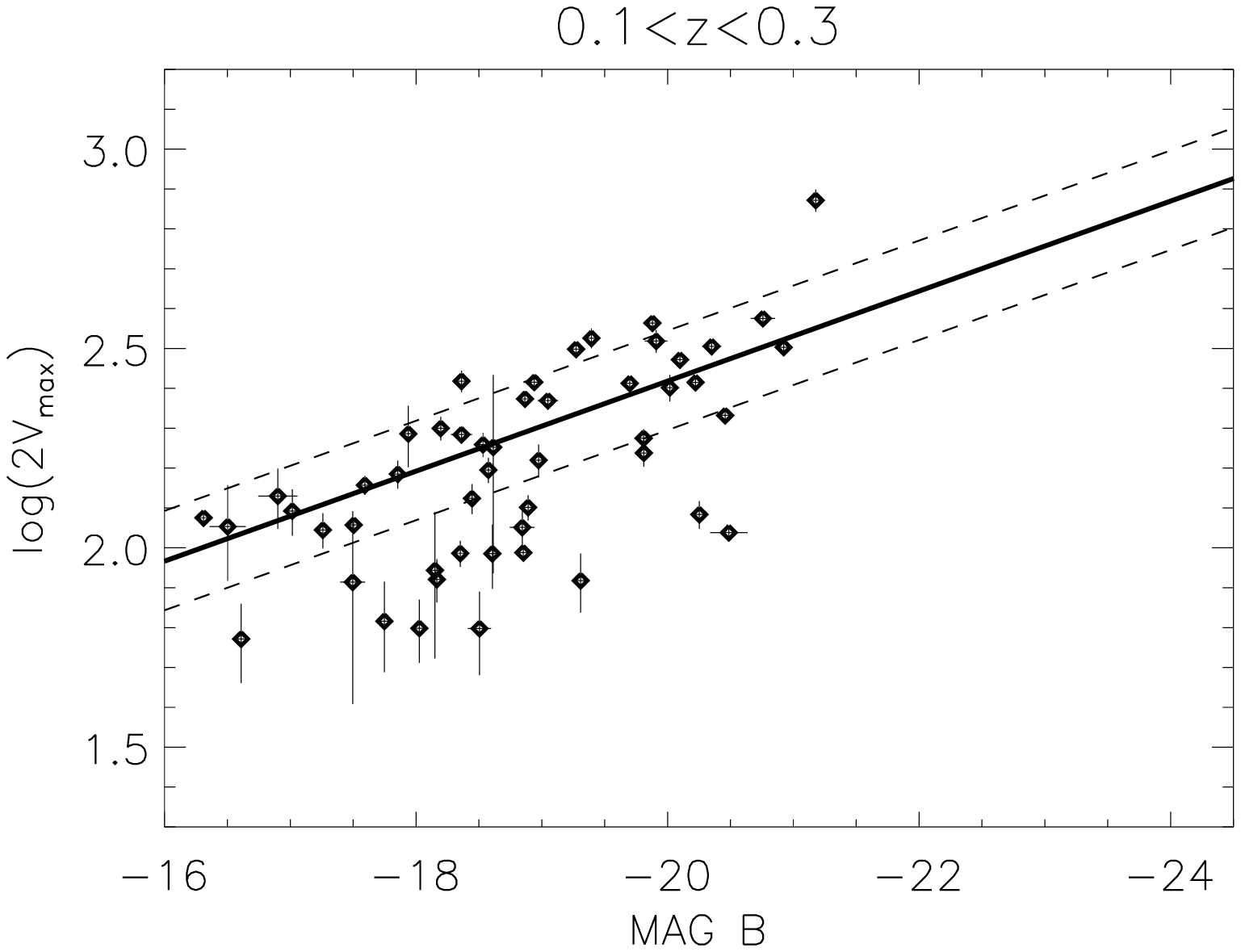}
\includegraphics[scale=0.46]{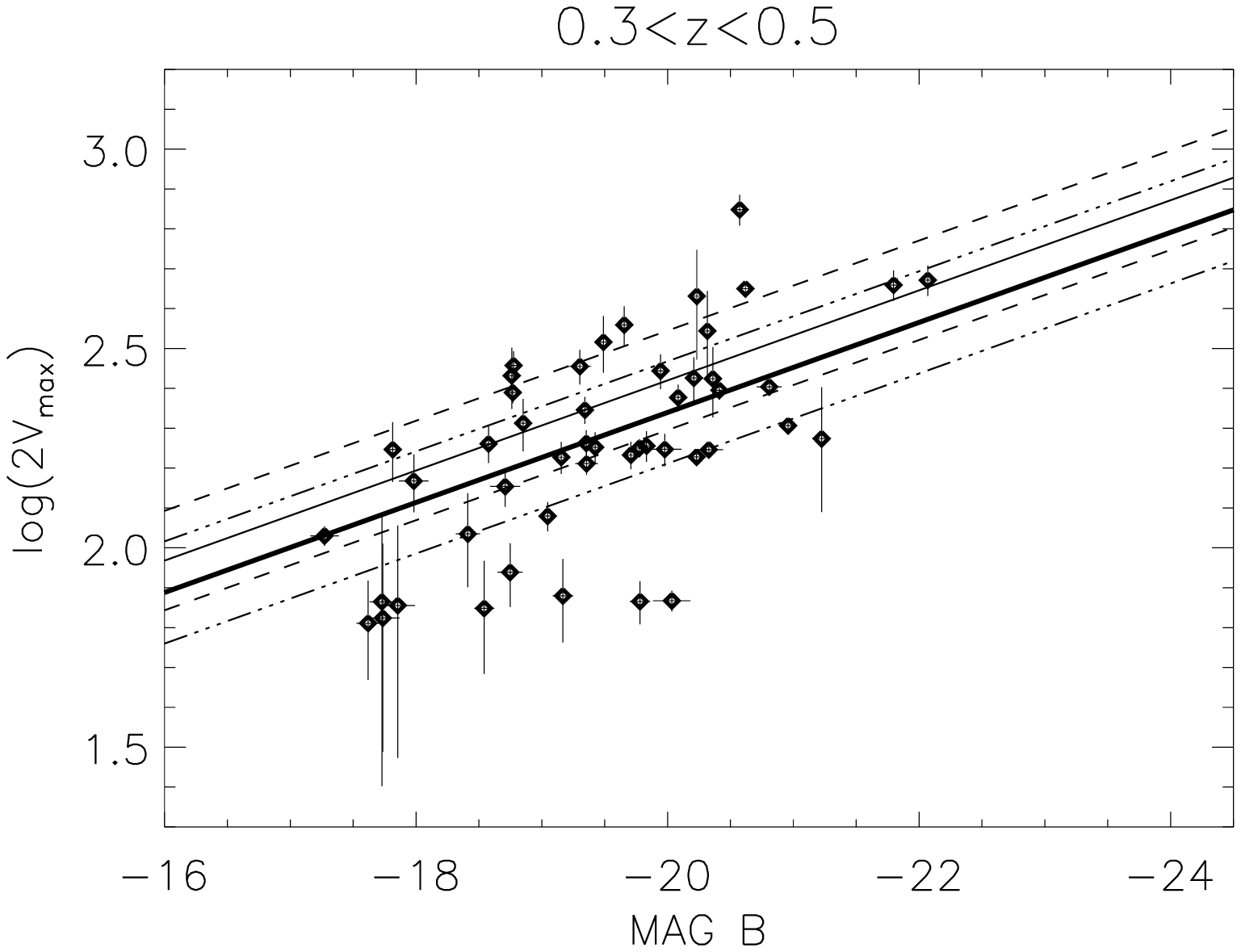}
\includegraphics[scale=0.46]{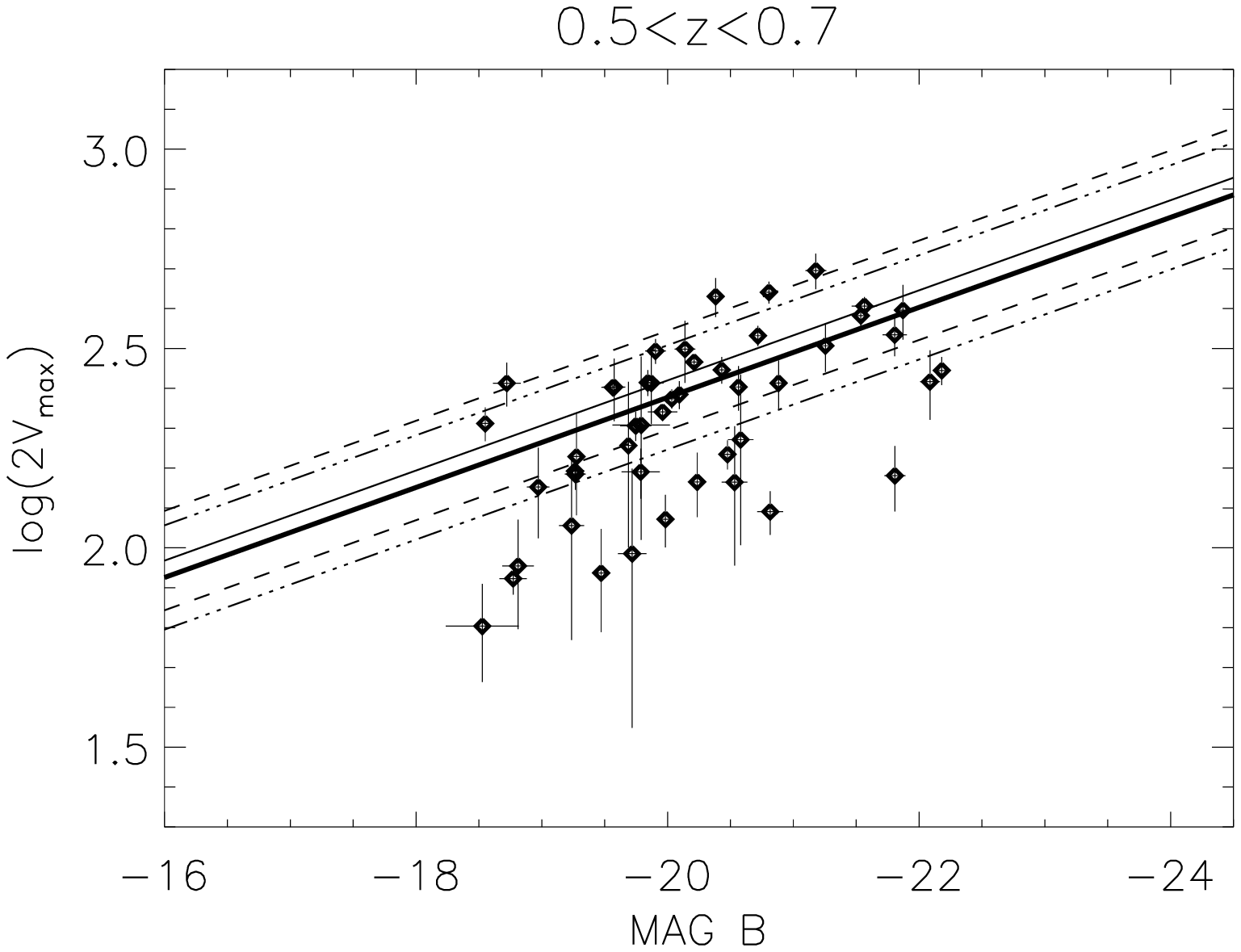}
\includegraphics[scale=0.46]{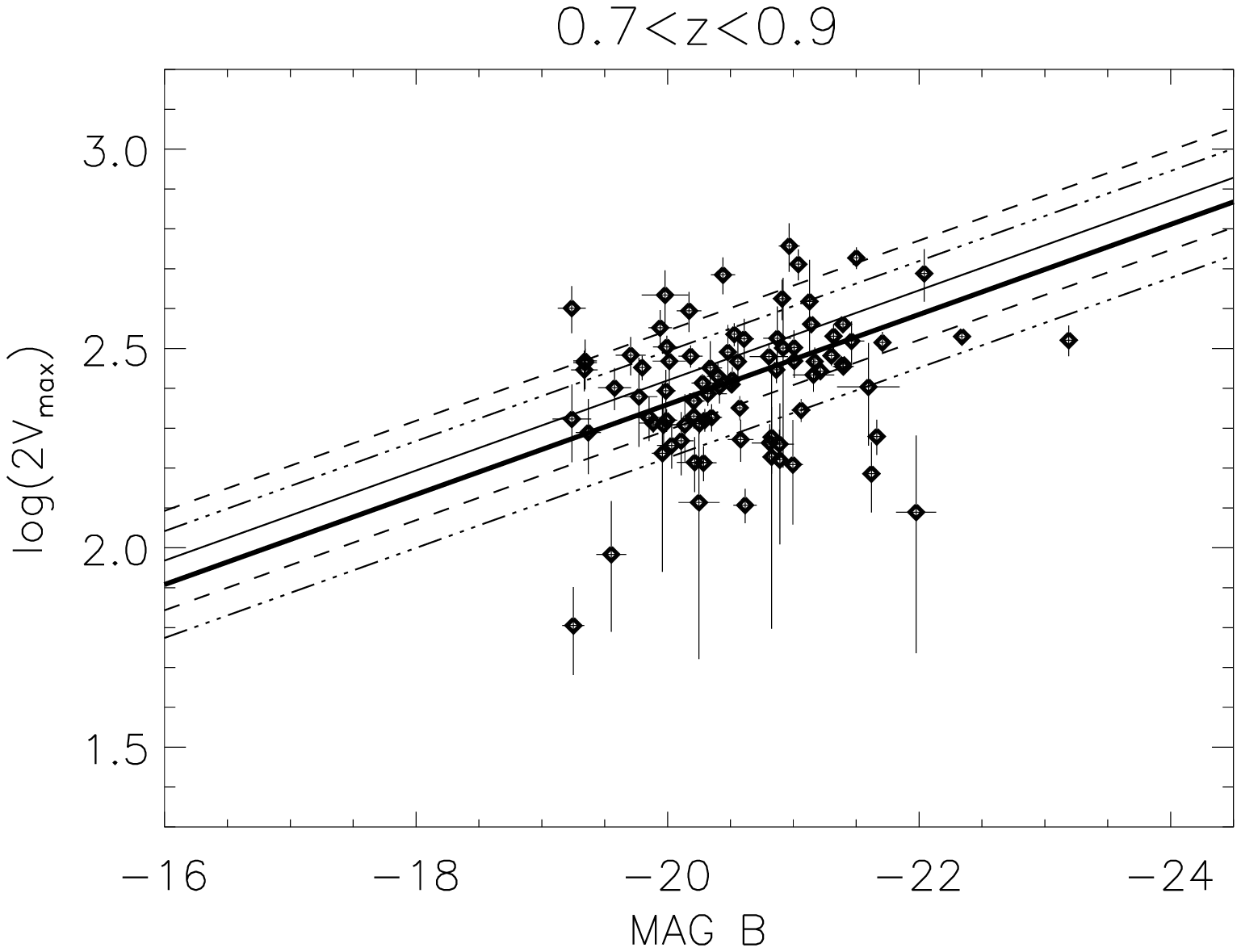}
\includegraphics[scale=0.46]{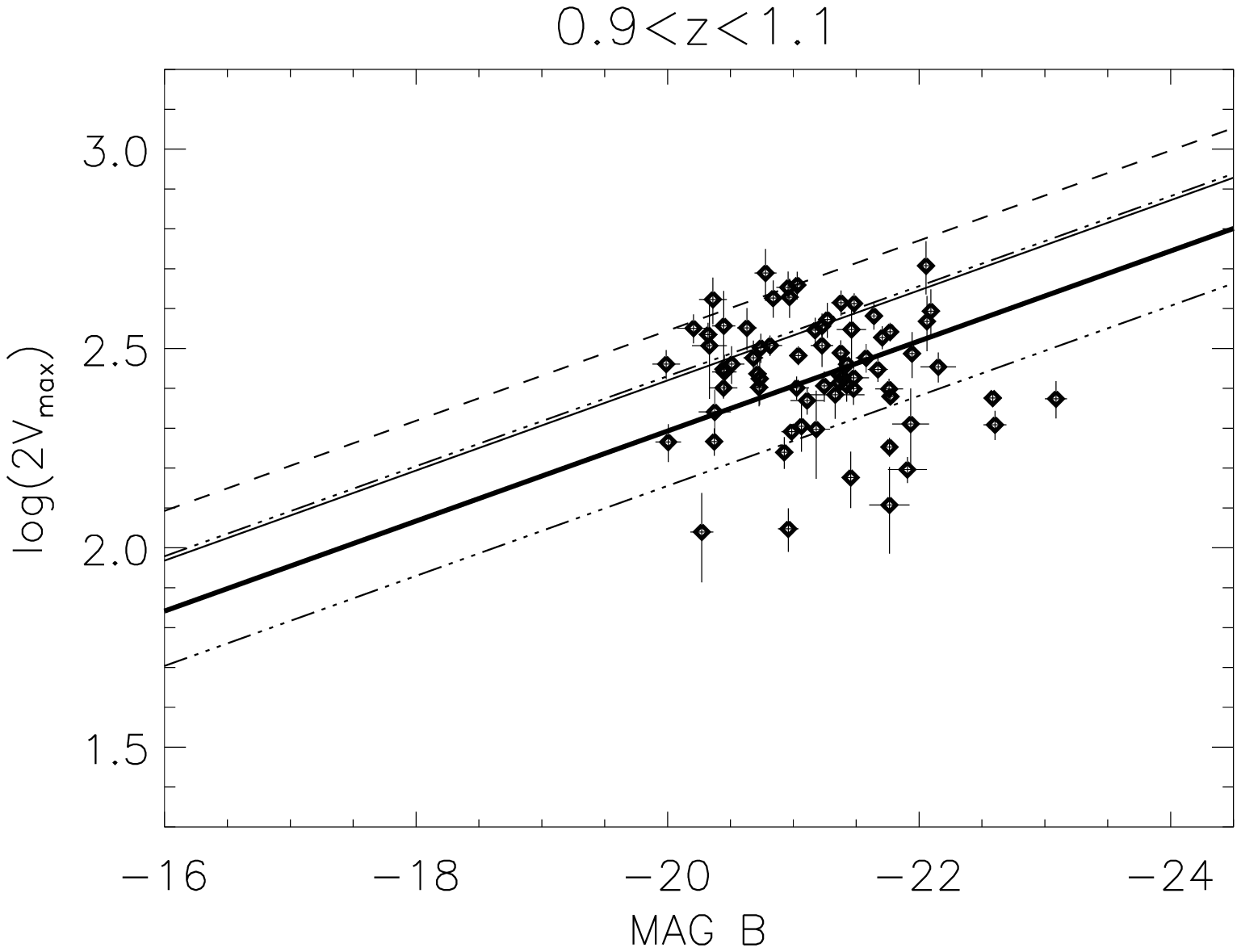}
\includegraphics[scale=0.46]{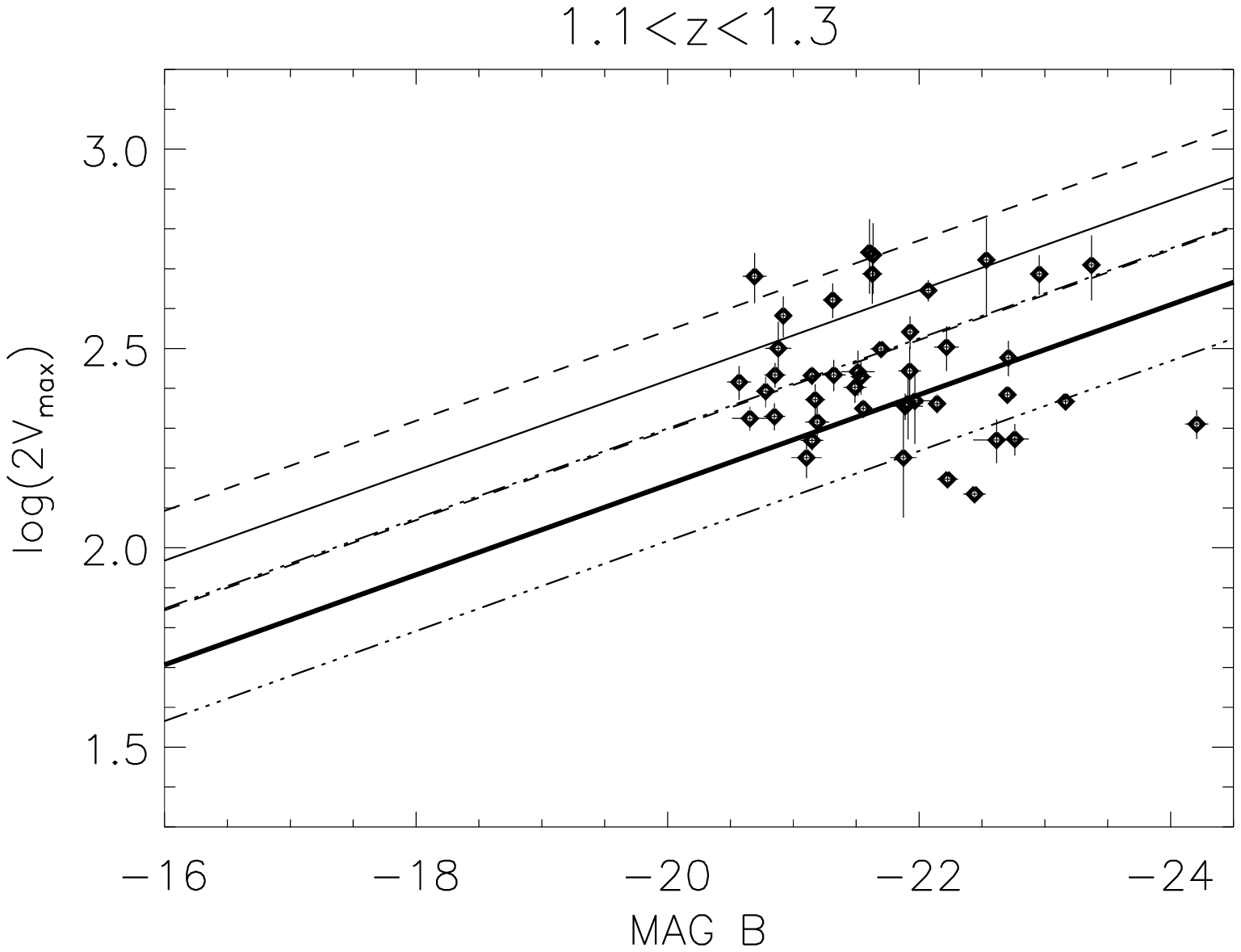}
\caption{Tully--Fisher relation in the B--band for 6 redshift ranges from 0.1 to 1.3. Thin and dashed lines represent, respectively, the local TFR and its 2${\sigma}$ uncertainty in the offset. The thick line is the weighted least--squares fit to the data, and the three dot--dashed lines are its 2${\sigma}$ uncertainty in the offset.}
\end{figure*}

As can be seen in Fig. 6, the zero--point of the TFR changes with redshift, since the difference with the local zero--point is larger as the redshift increases. For the last bin, corresponding to 1.1$<$z$<$1.3, the difference is more than 2${\sigma}$. It is important to note that the same zero--point change can be seen in both R and I--bands. A similar result for the lower redshift bins was found by \citet{2007IAUS..235....3V}, where all fits had lower zero--points than the local TFR. Nevertheless, in her work, all fits were within 2${\sigma}$, being 0.9$<$z$<$1.1 the last redshift range. She concluded that no TFR evolution was detected and that the change of the zero--point was instead attributed to objects with high equivalent widths. In Fig. 6, our result is the same up to $z=1.1$, but in the next redshift range (1.1$<$z$<$1.3), not only is the zero--point larger than 2${\sigma}$ uncertainty in the offset, but most galaxies ($\sim 85\%$) have their continuum brighter than their local counterparts. Then, it is unlikely that this zero--point change can be claimed as due to emitters of high equivalent widths. Nevertheless, and given the data scatter, a clear--cut statement on the evolution of the TFR up to z$=$1.3 requires higher redshift data, or data with lower associated errors. This type of work can be carried out using the new generation of 30 meter--class telescopes or with the deeper observations of DEEP3.

\begin{figure}
   \centering
   \includegraphics[scale=0.46]{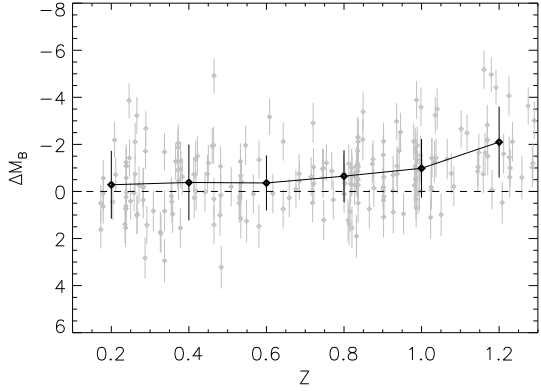}
      \caption{Magnitude residuals of the DEEP2 sample as a function of redshift in the B--band. Only the galaxies with a velocity error lower than 5\% have been represented, so that the figure is visually clear. The black points are the weighted average in each redshift range represented in Fig. 6.}
         \label{Fig7}
   \end{figure}

To quantify the changes in the intercept of the TFR, it is helpful to work with the magnitude residuals resulting from the relation, that is, the difference between the absolute magnitude of one galaxy and the magnitude from the local TFR (for its velocity). In Fig. 7, these residuals versus redshift are represented in the B--band. A tendency for galaxies to be brighter in the past can be seen, although at the level of the relation scatter. In addition, we represented the weighted average to the data in the same redshift ranges as in Fig.6. No change in magnitude for z $\leq$ 0.6 can be seen. Then, the galaxies become brigher as redshift increases.

Although the data could be fit by a non--linear function, since data dispersion is quite high, it is difficult to determine whether or not a non--linear fitting can really provide a more realistic approximation, so we have chosen a linear fit, which provides a qualitative idea of the evolution found. The least--squares fit to the data in each band, weighted by the errors in ${\Delta}M_Q$ and obtained through error propagation, yields the following relations:

\begin{equation}
{\Delta}M_B = (-0.13 \pm 0.12) \ + \ (-1.00 \pm 0.15) \ z
\end{equation}

\begin{equation}
{\Delta}M_R = (0.00 \pm 0.12) \ + \ (-1.00 \pm 0.15) \ z
\end{equation}

\begin{equation}
{\Delta}M_I = (0.04 \pm 0.12) \ + \ (-1.01 \pm 0.16) \ z \ .
\end{equation}
However, this procedure is only valid if the TFR slope does not change with redshift, because the magnitude range is not the same at low as at high redshift.

In the three bands, the results are very similar, with a change in magnitude at z=1, ${\Delta}$M$=-1.1{\pm}0.2$ in B and I--bands, and ${\Delta}$M$=-1.0{\pm}0.2$ in R--band. This result is in agreement of that found by \citet{2006MNRAS.366..308B} and \citet{2004A&A...420...97B}, although these studies were carried out using only the B--band. In a more recent work by \citet{2007ApJ...668..846B}, a luminosity evolution of $(-1.22\pm0.40)$z was found for galaxies with redshift $<z>=0.5$. Taking the errors into account, this result would agree with our work, too. However, these authors claim that this difference is found only for low--mass galaxies, whereas the distant high--mass spirals are compatible with the local Tully--Fisher relation. In our case, the high--mass galaxies are brighter than local galaxies, in contrast to \citet{2007ApJ...668..846B} work.

\section{Discussion: Tully--Fisher evolution and its causes}

A change in the TFR with lookback time represents a variation in luminosity, velocity, or both. A velocity change can happen if the mass of the galaxy disc changes with redshift, for a fixed luminosity. The mass can increase by galaxy merger processes, but this would only be noticeable if the mass of the merger is at least $\sim$15\% of the parent galaxy, whereas in this case the merger would destroy the galaxy disc. \citet{2007MNRAS.375..913P} discuss the TFR evolution as predicted by cosmological simulations of disc galaxy formation, including hydrodynamics, and all the relevant baryonic physics (star formation, chemical and photometric evolution, metal--dependent cooling, energy feedback). These simulations predict a growth of the disc mass by infall of halo gas and ongoing star formation. This would increase the rotation velocity of the galaxy, but at the price of increasing galaxy luminosity as well. Then, in this model, the evolution would shift the position of the galaxies along the TFR. Then, a change in luminosity seems more likely than a velocity change. Within this scenario, assuming pure luminosity evolution, a change in the slope of the TFR would mean a different evolution for high and low--mass galaxies, while a zero--point TFR change would indicate a uniform evolution. 

The \citet{2007MNRAS.375..913P} simulations show a B--band luminosity evolution ${\Delta}M_B=-0.85$ magnitudes at z=1, with a constant TFR slope. This result is less than the evolution found in this work, although in the same direction: galaxies would have been brighter in the past, for a fixed velocity. Their models predict a luminosity fading and reddening caused by the ageing in the stellar populations, combined with the increase of the stellar mass and rotation velocity as stated above. One point in favour of this theory is the increase in the star formation rate (SFR) density from z=0 up to z=1--2 observed in numerous works \citep[see, for example,] []{1997ApJ...489..559G,1998ApJ...498..106M,2005ApJ...633L...9F}. Although the mass locked in stars would still actually be growing, their growth would have been greater in the past, affected by stellar populations ageing. Were this explanation correct, the effect should be more noticeable in blue--bands. However, in this work, the same change in luminosity in B, R, and I--bands has been found. 

The local TFR is colour dependent, in other words, the galaxies that rotate faster are redder. In Fig. 8 we represent the colour R--I of our galaxies versus the rotation velocity and the local difference between R and I--band TFRs. For the three redshift ranges, all galaxies follow the colour of the local relation, although with larger scatter for higher redshift galaxies. Then, the colour of a galaxy for a fixed velocity does not change significantly with redshift. Also, were the slope of the TFR constant with redshift as other works suggest \citep{2006MNRAS.366..308B}, the colour of the Tully--Fisher relation would not change with the redshift.

Finally, \citet{2002A&A...389..761W} studied the colour evolution of disc galaxies in two models of galaxy formation: the accretion model, and the collapse model. In both models, the intrinsic integrated colour evolution R--I would be lower than 0.2 in the redshift range 0$<$z$<$1.2, within the dispersion shown in Fig 8. However, the change in luminosity found in our work can only be reproduced by the collapse model. The greatest colour evolution difference between both models can be found in V--K. In this case, while the collapse model presents a change in colour lower than 0.1, the colour evolution in the accretion model at z=1.2 would be 0.6. Then, the study of optical--infrared colours in TFRs can provide insight into the mechanism of formation of disc galaxies.

\begin{figure*}
   \centering
   \includegraphics[scale=0.76]{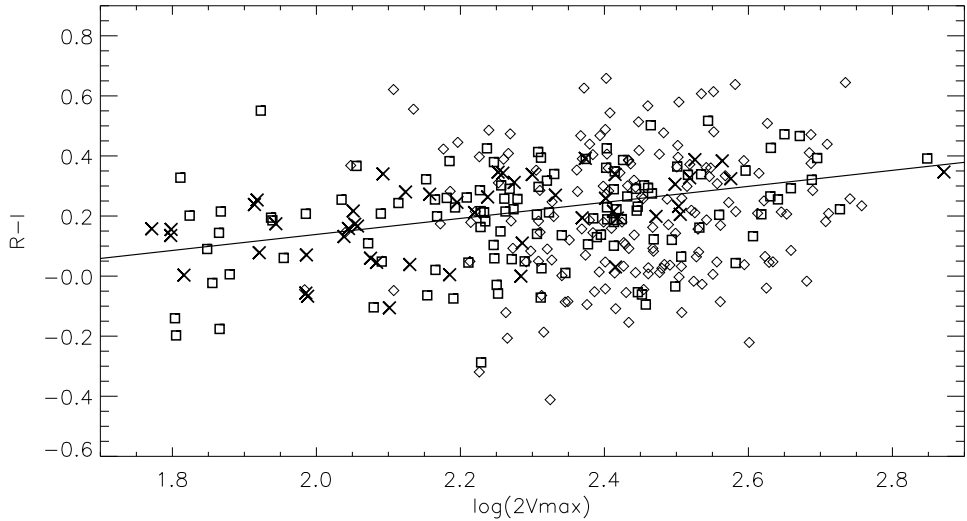}
      \caption{R--I absolute magnitude versus rotation velocity in the redshift range 0.1$<$z$<$1.3. Solid line shows the difference between local R and I band TFRs. The crosses indicate the galaxies in the redshift range 0.1$<$z$<$0.3. The squares are galaxies with 0.3$<$z$<$0.8, and the diamonds represent the objects with 0.8$<$z$<$1.3.}
         \label{Fig8}
   \end{figure*}

\section{Conclusions}

This study investigates the Tully--Fisher relation and its evolution with redshift in B, R, and I--bands. Using HST imaging data and DEEP2 spectroscopy taken with the Keck II telescope, galaxy characteristic parameters, such as rotation velocity and absolute magnitude, have been derived. The rotation velocity of a sample of 344 spiral galaxies in the Groth field were measured from optical lines widths. The objects span the redshift range 0.1$<$z$<$1.3. In this work we demonstrated that all optical emission lines can be used for determining disc--rotation velocities, if the signal--to--noise is high enough. Therefore, it is possible to extend the study of the TFR in the optical domain up to z=1.6.

In addition, we used three different techniques for morphological classification: comparison of the spectra with an elliptical/S0 template, visual classification, and GIM2D. The best method was the visual classification, and we found an error of $\sim$25\% in the selection, based on the S\'ersic index. 

Local Tully--Fisher relations were constructed from DEEP2 data in the redshift range 0.1$<$z$<$0.3, and were compared with \citet{2001ApJ...563..694V} relations. Despite the different ways of determining rotation velocity, the local relations derived in this work are consistent with his results, so our local TFRs have been adopted as low--redshift comparison patterns.

The B, R, and I--band TFRs were derived for 6 redshift ranges. A least--squares fit of the data in each range was obtained, maintaining the slope constant, for studying the evolution of the zeropoints. We find {\it prima facie} evidence of evolution of the TFR, with a tendency for galaxies to be brighter in the past. The same zero--point changes can be seen in R and I--bands. We suggest that this evolution is due to an ageing of the stellar populations as consequence of a star formation decrease down to z=0.  

The colour and luminosity evolutions found in our work support the collapse model rather than the accretion model of disc galaxy formation, showing that V--K colour evolution of TFRs can provide useful tools for studying galaxy formation.

\begin{acknowledgements}

This work was supported by the Spanish {\em Plan Nacional de Astronom\'ia y Astrof\'isica} under grant AYA2005--04149. We thank the DEEP2 group for making their catalogues and data publicly available and especially to Michael Cooper, who kindly helped us with the DEEP2 database. Funding for the DEEP2 survey has been provided by NSF grants AST95--09298, AST--0071048, AST--0071198, AST--0507428, and AST--0507483, as well as NASA LTSA grant NNG04GC89G. The work is based on observations obtained at the Canada-France-Hawaii Telescope (CFHT) which is operated by the National Research Council of Canada, the Institut National des Sciences de l'Univers of the Centre National de la Recherche Scientifique of France,  and the University of Hawaii.

This study makes use of data from AEGIS, a multiwavelength sky survey conducted with the Chandra, GALEX, Hubble, Keck, CFHT, MMT, Subaru, Palomar, Spitzer, VLA, and other telescopes and supported in part by the NSF, NASA, and the STFC.

We thank the SAO/NASA Astrophysics Data System (ADS) that is always so useful.

\end{acknowledgements}

\end{document}